\documentclass[12pt]{iopart}

\usepackage{iopams}
\usepackage{graphicx}
\begin{document}

\title[]{Uniqueness transition in noisy phase retrieval}

\author{Veit Elser$^1$ and Stefan Eisebitt$^2$}

\address{$^1$ Department of Physics,
Cornell University,
Ithaca, NY 14853, USA}

\address{$^2$ Institut f\"ur Optik und Atomare Physik, Technische Universit\"at Berlin, Hardenbergstr. 36, 10623 Berlin, Germany}

\ead{ve10@cornell.edu}

\begin{abstract}
Previous criteria for the feasibility of reconstructing phase information from intensity measurements, both in x-ray crystallography and more recently in coherent x-ray imaging, have been based on the Maxwell constraint counting principle. We propose a new criterion, based on Shannon's mutual information, that is better suited for noisy data or contrast that has strong priors not well modeled by continuous variables. A natural application is magnetic domain imaging, where the criterion for uniqueness in the reconstruction takes the form that the number of photons, per pixel of contrast in the image, exceeds a certain minimum. Detailed studies of a simple model show that the uniqueness transition is of the type exhibited by spin glasses.
\end{abstract}

\maketitle

\section{Introduction}

The x-ray phase problem, and inverse problems more generally, are often characterized as overdetermined or underdetermined \cite{Millane, Hauptman}. A successful, i.e. unique, phase reconstruction belongs to the former class, where the number of data exceed the number of free variables (atomic coordinates, contrast pixels, etc.). This way of formulating the feasibility criterion, however, may not be appropriate for any of several reasons. Intensity data is subject to noise, and in the extreme shot-noise limit can hardly be treated as a collection of continuous constraints. On the other side of the equation, the variables to be reconstructed may be quite different from simple real numbers. Examples are the binary-valued contrast in magnetic scattering from an Ising magnet, or the contrast of a set of identical atoms at low resolution. These examples highlight the fact that in many applications the actual information we wish to extract from the data is a small fraction of the information in a general image, and as a result the reconstruction should be able to tolerate significant noise in the data.

The question of uniqueness takes on a different character when we depart from the Maxwell constraint counting principle as applied to continuous contrast values and constraints. We will use the example of magnetic scattering to frame this question in precise terms. Suppose we are trying to reconstruct a 2D Ising domain pattern from its circular dichroism contrast in an x-ray scattering experiment \cite{experiment}. Given the x-ray wavelength  and the maximum scattering angle, the photons collected at the detector provide information about the domain pattern at a resolution of finitely many pixels, say $N$. Each pixel has one of two contrast values that we wish to determine. Further suppose that the total number of collected photons is a certain multiple of the number of pixels, say $\mu N$. Strong limits on the number of photons may arise in the case of stroboscopic single-shot experiments with x-rays, or simply due to weakness of the contrast. A basic question given these circumstances is whether some fraction of the pixels will always be uncertain for any $\mu$, or whether there is a critical $\mu_c$ above which this fraction is so small that a unique reconstruction, in a practical sense, is possible in principle.

\section{Communication channel analogy}\label{analogy}

The above question has a close analogy with problems studied in communication theory \cite{Shannon}. Consider a scheme where information, in the form of binary sequences, is transmitted on a noisy communication channel. The first stage in the process is to encode blocks of size $N$ by their Fourier intensities. Whereas in a diffraction experiment this is accomplished by the quantum mechanics of scattering, we will adhere to our information processing scenario where the intensities are obtained as the squared magnitudes of the discrete Fourier transform applied to the sequence of bits. Noise in our channel arises from the fact that it is not the continuous intensity values that are ``received" by the detector, but discrete photons. The final stage of our communication channel is thus the Poisson sampling of the Fourier intensities to give a sequence of $N$ non-negative photon counts. We specify the noise in the channel in terms of the mean number $\mu$ of photons per intensity sample. Examples of received signals (photon counts) for a particular binary input sequence, at various values of $\mu$, are shown in Table 1.

\renewcommand\arraystretch{1.5}
\begin{table}[!t]
\begin{center}
\begin{tabular}{c|*{13}{r}}
$\mu$&$+1$&$-1$&$+1$&$-1$&$-1$&$+1$&$-1$&$+1$&$+1$&$+1$&$+1$&$+1$&$-1$\\
\hline
5&4& 8& 2& 0& 2& 10& 6& 8& 10& 3& 0& 1& 9\\
10&8& 16& 3& 4& 0& 16& 9& 16& 31& 1& 3& 2& 18\\
15&5& 28& 2& 1& 4& 38& 21& 26& 38& 5& 3& 2& 21\\
\end{tabular}
\end{center}
\caption{Example of a binary-valued $\pm 1$ sequence (top row) and its encoding (lower rows) by Poisson samples (photon counts) of its Fourier transform intensity for three values of the mean photon number per bit, $\mu$. Our analysis shows that the original binary sequence can be reliably decoded from the Poisson sampled intensities, up to symmetries, when $\mu$ exceeds a value near 10.}
\end{table}

A central question addressed by communication theory is the extent to which the received signal in the noisy channel can be decoded to recover the original message. This hinges upon two things: (1) a capacity intrinsic to the characteristics of the channel, and (2) the encoding/decoding protocol. Whereas the first point also applies to the problem of uniqueness in imaging, the analogy breaks down somewhat when it comes to the second point. In the context of communications, Shannon \cite{Shannon} showed that information transmission approaching arbitrarily close to the rate given by the channel capacity is achievable by protocols that use carefully constructed codes. This flexibility in extracting the maximum information in a diffraction experiment is usually not available. For example, in the case of magnetic imaging the ``code" is already set by the form of the contrast: binary-valued pixels with some degree of correlation. The communications channel analogy thus applies less in the construction of capacity-achieving codes and more in the performance of given codes at various levels of noise.

There are trivial and exotic forms of non-uniqueness that qualify the binary-contrast decoding problem. These all apply even in the case of zero noise ($\mu\to \infty$). Trivial non-uniqueness arises as a result of symmetry. Cyclically shifting a binary sequence, reflecting it, or reversing the contrast values (when these are $\pm 1$), all do not change the Fourier intensities. This follows from the fact that these operations preserve the sequence-autocorrelation, which by a Fourier transform is equivalent to the intensities. Exotic non-uniqueness can be understood by the same device. Suppose a binary sequence $b$ factors as the cyclic convolution product of two binary sequences: $b=b_1\ast b_2$. The convolution $b'=b_1\ast R(b_2)$, of $b_1$ with the reversal of $b_2$, will then have the same cyclic autocorrelation, and therefore intensity, as $b$. Consequently if $b'$ is binary and not related by a symmetry to $b$, then we have at least two decodings. The two sequences of length $N=13$ below are an example of this phenomenon:
\begin{equation*}
\begin{array}{rrrrrrrrrrrrr}
+1& +1 &+1& +1& +1 &-1& +1& -1 &-1& +1 &+1 &+1 &-1,\\
+1& +1& +1& +1& +1& +1& -1& +1& -1& +1& +1& -1& -1.
\end{array}
\end{equation*}
We could dispense with the above cases of non-uniqueness in the absence of noise by agreeing that the actual content of the ``messages" is the sequence-autocorrelation. For the most part this is the same as treating the magnetic contrasts as symmetry classes, the exotic cases being relatively rare\footnote{We believe the fraction of sequences which have the same autocorrelation as a sequence not in the same symmetry class vanishes exponentially with the sequence length $N$.}.

Error-correcting codes in communications exploit block structure to realize information rates approaching Shannon's channel capacity. With suitable encoding/decoding the probability of even one bit being flipped can be made to vanish exponentially in the size of the block when the information rate is below the channel capacity. We show below that a similar phenomenon appears to apply to diffractive imaging, where the ``encoding" is always the same (and not under the control of the imaging scientist). In concrete terms this means that the fidelity of magnetic contrast reconstruction is typically much better than naive estimates would predict.

\section{Capacity of the Poisson channel}

In communications theory the discrete-time Poisson channel is defined by the conditional probability (or transition matrix)
\begin{equation}\label{Poisson}
p(k|w)=\frac{w^{k}}{k!}\exp{(-w)},
\end{equation}
for receiving $k$ photons when the input intensity\footnote{By ``intensity" we actually mean photon fluence integrated over the area of a detector pixel, a dimensionless quantity.} signal is $w$. The capacity of the channel \cite{Shannon} is found by maximizing the mutual information associated with the joint probability
\begin{equation}
p(k,w) = p(w) p(k|w)
\end{equation}
with respect to the prior distribution $p(w)$. Mutual information is an information-theoretic measure of the degree of correlation of two random variables\footnote{We use capitalized variable names for all random variables.} and in this case is written $I(K,W)$.  The mutual information is the difference of two entropies
\begin{equation}
I(K,W)=H(K)-H(K|W)
\end{equation}
that, in our case, quantify the information provided by the photon counts $k\in K$ about the intensity $w\in W$, with due regard to the loss of information (second term) due to entropy in the counts even when $w$ is known precisely. The entropy and conditional entropy have the following forms when expressed in terms of the probability distributions:
\begin{eqnarray}\label{H(K|W)}
H(K|W)&=&-\sum_{k\in K}\sum_{w\in W}p(k,w)\,\log_2{\,p(k|w)}\\ \label{H(K)}
H(K)&=&-\sum_{k\in K}\sum_{w\in W}p(k,w)\,\log_2{ \sum_{w'\in W}p(k,w')}.
\end{eqnarray}

The maximization of $I(K,W)$ with respect to $p(w)$, to determine the channel capacity, is usually performed with some constraints on $p(w)$ that fix the mean or maximum value of $W$. Since one has very little control over the distribution $p(w)$ in diffractive imaging, other than its mean value, it makes sense to define the capacity of the Poisson channel for the particular form of prior $p(w)$ that applies in most cases. In diffraction theory, the form
\begin{equation}\label{Wilson}
p_\mu(w)=\exp{(-w/\mu)}/\mu
\end{equation}
is known as Wilson statistics \cite{Wilson} and arises when the complex-valued radiation amplitude has a Gaussian distribution as it would when the contrast pixels are modeled as independent random variables. Here and below, $\mu$ is the mean number of photons associated with one intensity measurement.

The information capacity of the Poisson channel, for the prior distribution (\ref{Wilson}), can be calculated using the expressions (\ref{H(K|W)},\ref{H(K)}) and takes the following form \cite{PCC},
\begin{equation}\label{IP}
I_P(\mu)=(\mu+1)\log_2{(\mu+1)}-\frac{\gamma \mu}{\log{2}} -\sum_{k=2}^\infty \frac{ \log_2{k}}{(1+\frac{1}{\mu})^k},
\end{equation}
where $\gamma$ is Euler's constant. We are not aware of any work that considers the construction of block codes that achieve this information capacity. A block code of length $N$ in this context would be a collection of blocks of intensities $(w_0,\ldots,w_{N-1})$ with average value $\mu$. We show below that block codes arise naturally in the context of diffractive imaging and it is their error correcting capacity that reconstruction algorithms can take advantage of.

\section{The Fourier-Poisson channel}

For diffractive imaging it makes sense to define the ``communication channel" in a way that takes into account the block structure of the signal. We will continue our discussion for the case of signals in one dimension, as introduced in section \ref{analogy}, but the same construction applies to imaging in two and three dimensions.

For our signals we will take vectors of contrast values $x=(x_0,\ldots,x_{N-1})$ subject to some prior distribution $p(x)$. In many applications the contrast is real-valued and that will be the case we consider here. The first stage of our channel encodes the contrast into a vector of Fourier intensities $w=(w_0,\ldots,w_{N-1})$, where
\begin{equation}\label{intens}
w_n(x)=\frac{\mu}{N}\left|\sum_{m=0}^{N-1}e^{i \frac{2\pi}{N} m n} \,x_m\right|^2.
\end{equation}
The normalization provided by $\mu$ is such that if the contrast prior $p(x)$ satisfies the constraint $\langle x\cdot x\rangle=N$ on the average power, then the average intensity satisfies $\sum_{n=0}^{N-1} \langle w_n\rangle = N\mu$. The final stage of our channel is the Poisson sampling of the intensities with (\ref{Poisson}) to give the vector of photon counts (nonnegative integers) $k=(k_0,\ldots,k_{N-1})$.

We will refer to the combined operations of Fourier intensity encoding followed by Poisson sampling as the Fourier-Poisson channel, or FPC. The pair of channel variables are $X$, representing the contrast in the imaging experiment and its probability distribution, and $K$, 
the photon counts recorded by the detector. The associated mutual information, $I(X,K)$, is the information acquired in a typical experiment about the contrast in a typical sample. Our choice of constructing the ``block code" by the discrete Fourier transform in one dimension corresponds to imaging a one dimensional crystal. The $N$ intensities $w_0(x),\ldots,w_{N-1}(x)$ are the set of Bragg intensities extending to a resolution where the contrast $x=(x_0,\ldots,x_{N-1})$ is sampled at $N$ equally spaced points within the one dimensional unit cell. Because $x$ is real, the intensities have the Friedel symmetry property\footnote{In all our expressions indices are given up to an (irrelevant) multiple of $N$.}: $w_n(x)=w_{-n}(x)$.

In the case of the Gaussian prior
\begin{equation}\label{Gaussian}
p(x) = (2\pi)^{-N/2} \exp{(-{\textstyle \frac{1}{2}}\, x\cdot x)}
\end{equation}
the mutual information of the FPC can be directly related to the capacity of the simple Poisson channel (\ref{IP}). The discrete Fourier transform of $x$, written $\hat{x}$, is a linear transformation of $x$ with the property $\hat{x}\cdot\hat{x}^\ast=x\cdot x$. We may therefore work instead with the Gaussian distribution on the complex variables $\hat{x}_0,\ldots,\hat{x}_{N-1}$. These variables are an equivalent representation provided one takes into consideration the symmetry property $\hat{x}_{-n}=\hat{x}_{n}^\ast$. Since the intensities (\ref{intens}) are given by $w_n=\mu |\hat{x}_n|^2$, the FPC mutual information depends on only $\lfloor N/2+1\rfloor\approx N/2$ independent intensity distributions. With the exception of $w_0$ and $w_{N/2}$ (when $N$ is even), these are identical Wilson distributions (\ref{Wilson}). The exceptions are not of the form (\ref{Wilson}) because the corresponding $\hat{x}$ is real; however, their contribution to the mutual information in the large $N$ limit can be neglected.

\begin{figure}[t]
\begin{center}
\includegraphics{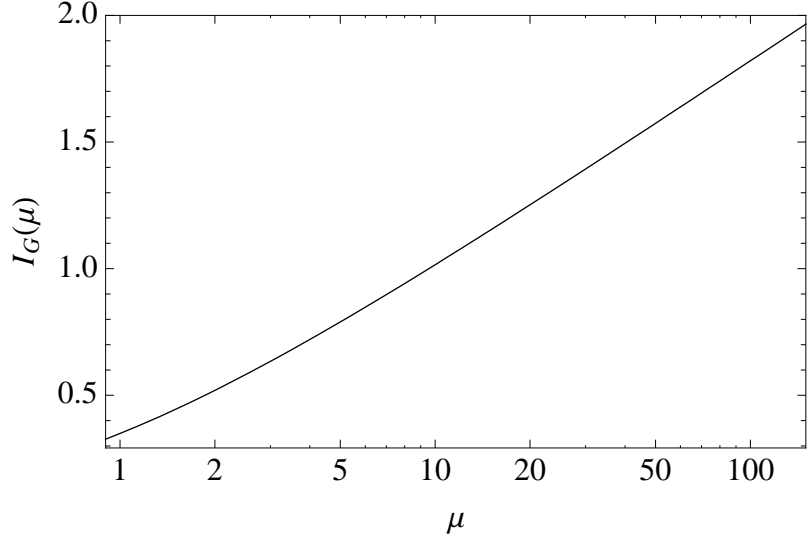}
\end{center}
\caption{Mutual information of the Fourier-Poisson channel as a function of the mean photon number $\mu$ when the contrast of the crystal has a Gaussian prior. The function $I_G(\mu)$ gives the number of bits of information available per Bragg peak when the average peak has $\mu$ photons.}
\end{figure}

In the preceding we showed that the mutual information for the Gaussian prior is, for large $N$, the mutual information associated with $N/2$ independent and identical joint distributions, a particular one involving $k_n$, $k_{-n}$ and $w_n$ (which always equals $w_{-n}$). The value of $I(X,K)$ for large $N$ is therefore given by $N/2$ times the mutual information associated with the joint distribution
\[
p(k_n,k_{-n}, w_n)=p(k_n|w_n)\, p(k_{-n}|w_n)\, p_\mu(w_n).
\]
Expressing this in terms of the variable $k_n^+=k_n+ k_{-n}$ (instead of $k_{-n}$), the distribution takes the form
\[
p(k_n^+,k_n,w_n)=\frac{w_n^{k_n^+}}{k_n! (k_n^+-k_n)!}\exp{(-2 w_n)}\,p_\mu(w_n)
\]
which is the product of a function of $k_n$ and a function of $w_n$. The mutual information for this pair of variables therefore vanishes and thus the mutual information is the same as that for the marginal distribution
\[
p(k_n^+, w_n) = \sum_{k_n=0}^{k_n^+}p(k_n^+,k_n,w_n)=\frac{(2 w_n)^{k_n^+}}{k_n^+!}\exp{(-2 w_n)}\,p_\mu(w_n).
\]
Expressing this in terms of the variable $w_n^+=2 w_n$, we obtain
\[
p(k_n^+, w_n^+) = \frac{(w_n^+)^{k_n^+}}{k_n^+!}\exp{(-w_n^+)}\,p_{2 \mu}(w_n^+),
\]
which is the same as the probability distribution that defines the simple Poisson channel but with $\mu$ replaced by $2\mu$. We therefore obtain the following result for the mutual information per contrast element of the FPC with Gaussian contrast prior:
\begin{eqnarray}
I_G(\mu) &=& \lim_{N\to\infty}I(X,K)/N\\
&=&\frac{1}{2}I_P(2\mu).
\end{eqnarray}

The function $I_G(\mu)$ is plotted in Figure 1. Although it was derived for the case of a one dimensional crystal, the same result holds for periodic contrast in any number of dimensions.  In concrete terms, this mutual information measures the maximum number of bits of information per Bragg peak we can obtain about the contrast in a crystal as a function of the mean number $\mu$ of photons detected in a typical Bragg peak. For example, suppose we have a weakly scattering crystal and collect only 100 photons per Bragg peak on average. In this case the information in $M$ Bragg peaks will provide about 2 bits of information about the contrast (e.g. electron density) at $M$ sample points in the unit cell. In the remainder of this paper we will explore the implications of this function for a type of contrast where even 1 bit of information is sufficient to reconstruct the essential structure of the sample.

\section{Fidelity of binary contrast reconstruction}\label{Fidelity}

We now turn to the problem of reconstructing a signal $x$ known to be binary, that is, where each component takes only the values $\pm 1$. In this case the prior probability $p(x)$ is the uniform distribution on the hypercube $B=\{-1,1\}^N$. Alternatively, we can continue to work with the FPC defined by the Gaussian prior and view the set $B$ as a code. 
From the capacity $I_G(\mu)$ per contrast element of the FPC we derived in the previous section we obtain the threshold value $\mu_c\approx 9.543$ where the capacity exceeds 1 bit. The most optimistic result for binary contrast reconstruction would therefore be that the code $B$ achieves exactly this capacity for $\mu>\mu_c$. We will see that although $B$ fails to achieve the channel capacity in a strict sense, something close to this does indeed hold.

To investigate reconstruction fidelity, that is, the error correction properties of our chosen binary code of signals, we consider the behavior of the optimum decoder. By definition, the latter is given by the maximum likelihood principle, where the decoded $x\in B$ maximizes the conditional probability 
\begin{equation}\label{p(k|x)}
p(k|x)=\prod_{n=0}^{N-1}\frac{w_n(x)^{k_n}}{k_n!}\exp{(-w_n(x))}
\end{equation}
for any given received vector of photon counts $k$. We can simplify this decoding function by taking its logarithm and summing pairs of photon counts $k_n^+=k_n+ k_{-n}$ that arise from equal intensities:
\begin{equation}
d(k^+|x)=\sum_{n=1}^{\lceil N/2\rceil-1}k^+_n \log{w_n(x)}.
\end{equation}
For simplicity we have omitted from the sum the terms $n=0$ and $n=N/2$ (for even $N$) which are unpaired and can be neglected in the limit of large $N$, and henceforth abbreviate the index of the final term as $N/2$. We have also omitted the factorials of the photon counts since they are irrelevant when comparing the codewords $x$. Similarly, since
\[
\sum_{n=0}^{N-1} w_n(x) = \mu N
\]
for all $x\in B$, it too can be omitted in the definition of the decoding function.

\begin{figure}[t]
\begin{center}
\includegraphics{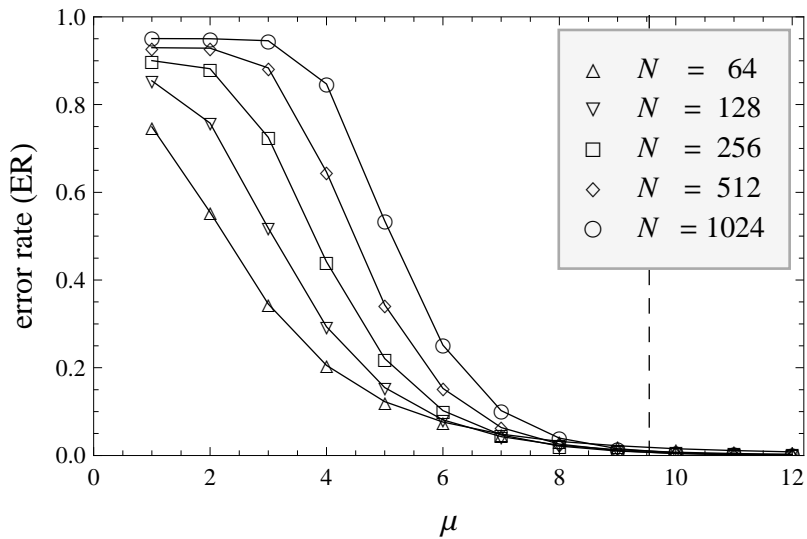}
\end{center}
\caption{Error rate (ER) in binary contrast decoding as a function of the mean photon number $\mu$ for block codes up to size $N=1024$. Each curve gives the probability that the contrast sampled at $N$ points within the unit cell of a crystal will be reconstructed with one flipped value when the best possible (maximum likelihood) reconstruction criterion is used. The dashed line marks the value $\mu=\mu_c$ where the Fourier-Poisson channel has capacity 1 bit.}
\end{figure}

The severest challenge for the decoding function is its ability to detect a single flipped bit. This follows from the fact that the Fourier transform, as a unitary transformation, preserves the Euclidean distance between codewords. A pair of codewords $x$ and $x'$ differing in just one bit (one sign reversal) have $2=\|x-x'\|=\|\hat{x}-\hat{x}'\|$, that is, Fourier transforms with the smallest possible separation. This is a sufficient, though not necessary condition for the corresponding intensities $w(x)$ and $w(x')$ to be close and difficult to discriminate. A pair of Fourier transforms might have a larger separation and yet have nearly the same intensities if all the complex amplitudes were mostly phase rotations. But apart from symmetry related codewords this scenario is statistically unlikely.

Our fidelity tests have therefore focused on the ability to detect a single flipped bit. We performed numerical simulations where a random codeword $x\in B$ is selected and the corresponding intensity vector $w(x)$ is Poisson sampled, and pairwise summed, to give a vector of photon counts $k^+$. The decoding function $d(k^+|x)$ is then evaluated and compared with $d(k^+|x')$, where $x'$ differs from $x$ by a single flipped bit. A decoding error is declared if $d(k^+|x')>d(k^+|x)$ for any of the $N$ bits that can be flipped. Finally, the decoding error rate, or ER, is obtained from the frequency of decoding errors when this procedure is applied to a million codewords $x$ selected uniformly from $B$.

Results of our fidelity tests are shown in Figure 2 and appear consistent with there being a  transition to unique reconstructions at a critical mean photon number $\mu_c$ near 10. For example, the small ER at $\mu=10$ for size $N=1024$ shows that with as few as ten photons per Bragg peak it is very unlikely that an optimal reconstruction will make an error in even one flipped contrast element out of all $1024$ in the unit cell. The Figure also shows that the ER tail extends beyond the analytically determined value $\mu_c\approx 9.543$ where the capacity of the FPC is 1 bit. To properly decide whether the binary block code achieves this capacity it is necessary to investigate the behavior of the ER in this tail, in particular, whether the ER vanishes in the limit of large $N$ for $\mu>\mu_c$.

\begin{figure}[t]
\begin{center}
\includegraphics{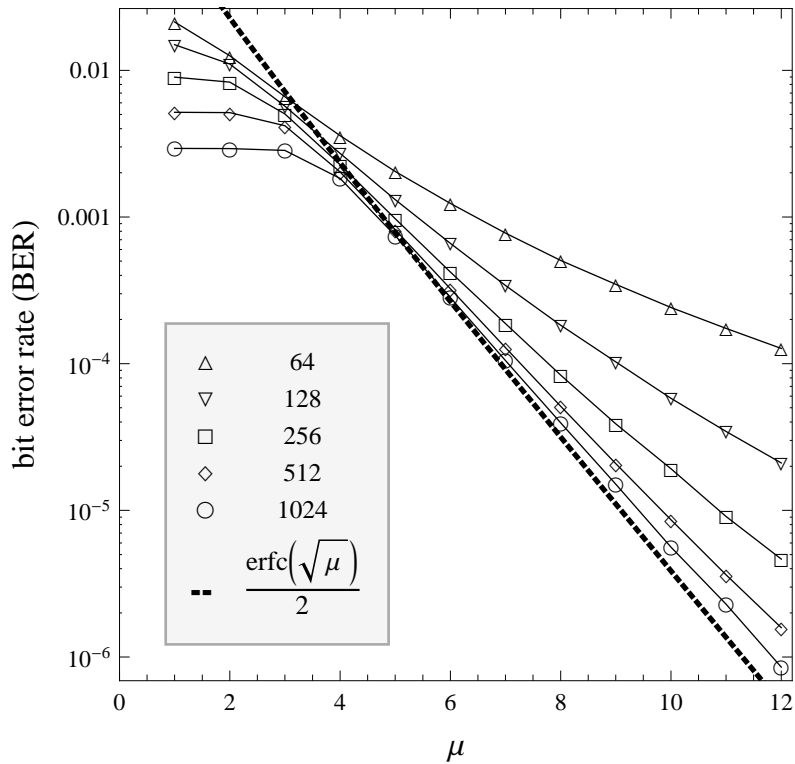}
\end{center}
\caption{Bit error rate (BER) corresponding to the data shown in Figure 2. For fixed mean photon number $\mu$, even in the region $\mu>\mu_c$, the BER does not vanish with the block size $N$ and instead approaches a limiting value given by the error function (dashed curve).}
\end{figure} 

To study the more rigorous decoding criterion we consider a quantity related to the ER that has a nicer limiting behavior with the block size $N$. This is the bit error rate, or BER, defined by
\begin{equation}
\mathrm{BER}=1-(1-\mathrm{ER})^{1/N}.
\end{equation}
Solving for ER in terms of BER we see that the latter corresponds to an interpretation of the ER as arising from independent bit-wise errors throughout the block. Figure 3 shows the BER obtained from the same data of Figure 2 on a logarithmic scale emphasizing the tail region. The convergence of the tail to a limiting form shows that the binary code does not achieve the full capacity of the FPC in the limit of large block size --- the rate of decoding errors remains finite for $\mu>\mu_c$.

The slight failure of the binary contrast code to achieve the 1 bit capacity of the FPC for $\mu>\mu_c$ may be more of interest to coding theory than the reconstruction problem in diffractive imaging. In any case, it is interesting to see that the BER vanishes apparently exponentially\footnote{A naive estimate based on the signal-to-noise ratio $\sqrt{\mu}$ in each of multiple independent intensity measurements would predict a much higher error rate.} in $\mu$ and it is this behavior that we now turn to.

Consider the random variable defined by
\begin{eqnarray}
\Delta(k^+,x)&=&d(k^+|x)-d(k^+|x')\\
&=&\sum_{n=1}^{N/2}k_n^+\log{\left(w_n(x)/w_n(x')\right)},\label{Delta}
\end{eqnarray}
where as before $x'$ is a single bit-flip applied to $x$. The vectors $x$ and $k^+$ are themselves random variables: $x$ has a uniform distribution on the set of binary codewords $B$ and $k_n^+$ is sampled from the Poisson distribution with mean $2 w_n(x)$. By symmetry the position of the flipped bit does not affect the distribution of $\Delta$ and we may take it to be the first bit. An event where $\Delta$ is negative corresponds to a decoding error since then the flipped vector $x'$ has higher likelihood than the vector $x$ from which the photon counts where sampled. If we can evaluate the probability of a negative $\Delta$ we will know the BER under the assumption that this quantity is dominated by single flip errors.

It is a straightforward exercise to determine the probability of a negative $\Delta$ if we can assume the distribution of this variable is Gaussian in the large $N$ limit. Both the mean and variance are found to have finite limits as $N\to\infty$:
\begin{equation}
\langle\Delta\rangle=4\mu\qquad\qquad
\langle\Delta^2\rangle-\langle\Delta\rangle^2=8\mu.
\end{equation}
Since these are the only parameters that determine a Gaussian, the BER is given as
\begin{eqnarray}
\mathrm{BER}&=&\frac{1}{\sqrt{2\pi}}\int_{\sqrt{2\mu}}^\infty \exp{(-t^2/2)}\, dt\\
&=&\frac{1}{2}\mathrm{erfc}(\sqrt{\mu})\\
&\sim&\exp{(-\mu)}/\sqrt{4\pi \mu}.
\end{eqnarray}
The exponential decay of the BER with $\mu$ is well supported by the simulations (Fig. 3). Although suboptimal in the sense of a code, for not vanishing for $\mu>\mu_c$, the rapid decay of the BER with photon counts has practical significance for diffractive imaging where the block size $N$ is fixed by the resolution and $\mu$ can be increased through the incident flux of radiation. The exponential behavior implies that substantial fidelity enhancement can be realized through rather modest increases of flux.

We conclude this section by speculating why Fourier intensity encoding of binary sequences apparently gives a very good, if not optimal, block code for the Poisson channel. Shannon \cite{Shannon} showed that a capacity achieving code, in the limit of large blocks, is realized by the construction where codewords are selected as random independent samples of the source distribution. In the case of the FPC with block size $N$ and $\mu>\mu_c$, where the capacity is 1 bit, this corresponds to drawing $2^N$ samples $x$ from the Gaussian (\ref{Gaussian}) and computing their intensities $w(x)$. Although this set of block-intensities is a capacity achieving code, it is not practical because the decoder needs the very long list of random $x$ used to generate the intensities. We believe the good behavior of the intensity code formed by just the binary $x$ is explained by the fact this set of block intensities is ``close" to being a random code as in Shannon's construction. Consider an arbitrary (not necessarily consecutive) selection of Fourier components $\hat{x}_1, \hat{x}_2, \ldots,\hat{x}_n$ with $n<N$ fixed. From the central limit theorem we know that these have independent complex-Gaussian distributions in the $N\to\infty$ limit even when $x$ is restricted to the binary contrasts $B$. The corresponding intensities $|\hat{x}_1|^2, |\hat{x}_2|^2, \ldots,|\hat{x}_n|^2$, in this fixed set of components, thus looks like a random code. The correlations that spoil the optimality of the binary construction are therefore rather global in scale, involving a number of Fourier components that grows with the size of the block. This reasoning also applies to our analysis above of the decoder variable $\Delta$ as it is a sum over Fourier components (\ref{Delta}). The independence properties of finite sets of these terms (as $N\to\infty$) lends support to the Gaussian statistics we assumed for their sum.

\section{Spin glass interpretation of the uniqueness transition}

The transition from a regime of image reconstructions with high rates of error to a regime of unique, practically error-free reconstructions, has an analog in statistical mechanics: the spin glass \cite{SpinGlass}. We explore this analogy for the case of binary contrast reconstruction but expect it to apply to phase retrieval problems more generally.

A spin glass is characterized by a paramagnetic phase at high temperature and multiple ``ordered" equilibrium phases at low temperature. In our reconstruction problem the role of temperature is played by shot noise and the control parameter is the mean photon number $\mu$. The thermodynamic phase behavior is exhibited not by any one particular reconstruction problem, but by the ensemble\footnote{It is in this respect that our treatment differs from other applications of spin glass ideas to the noisy channel coding problem \cite{Sourlas, MM}.} that comprises \textit{all} reconstruction problems of a particular size $N$. Either the algorithm (assumed to be optimal or near optimal) gives the correct reconstruction for almost all problems, or it almost always gets it wrong. This same ensemble appears in the definition of the mutual information $I_B(X,K)$ between binary codewords $X$ and their diffraction intensities as transmitted by the photon counts $K$. The distribution of the vector of photon counts
\begin{equation}\label{p(k)}
p(k)=\frac{1}{2^N}\sum_{x\in B}p(k|x).
\end{equation}
has the form of a sum of $2^N$ sub-distributions (\ref{p(k|x)}), grouped into $M$ symmetry classes (with respect to cyclic shifts, reflection, reversal) of binary codewords that have unique Fourier intensities. When the sub-distributions are well separated, as is required for a reconstruction algorithm (decoder) to have a low error rate, then the entropy $H(K)$ of the distribution $p(k)$ is simply the entropy $H(K|X)$ of a typical sub-distribution $p(k|x)$ augmented by the logarithm of the number of sub-distributions, or $\log_2{M}\sim N$. In this case we therefore have
\[
I_B(\mu)=\lim_{N\to\infty} I_B(X,K)/N=\lim_{N\to\infty} \log_2{M}/N=1
\]
for the mutual information per bit. On the other hand, at small $\mu$ or high noise, when the sub-distributions are not well separated, the entropy of $p(k)$ is close to the entropy of $p(k|x)$ for typical $x$ and the mutual information is small. The noisy reconstruction problem would exhibit a true thermodynamic phase transition in the large $N$ limit at some noise threshold $\mu_c$ if the sub-distributions became increasingly well separated with increasing $N$ for $\mu>\mu_c$.

In the previous section, however, it was shown that the decoding error rate for binary contrast remains finite (although very small) in the limit of large $N$. The sub-distributions $p(k|x)$ therefore do not become perfectly separated in the thermodynamic limit, and there is no phase transition in the strict sense. On the other hand, because the error rate vanishes so rapidly --- exponentially in $\mu$ --- the spin glass is still a useful analogy. The marginal distribution of photon counts (\ref{p(k)}) for the FPC is interesting as a definition of a (quasi) spin glass since the effective Hamiltonian\footnote{The effective Hamiltonian given by the logarithm of (\ref{p(k)}) is more complex in structure than standard spin glass Hamiltonians.}, when $p(k)$ is interpreted as a Boltzmann distribution, has no quenched disorder. Had we used $2^N$ random Gaussian samples (rather than simply all the binary samples) in the construction of our code for the FPC, then by Shannon's argument the error rate would vanish in the thermodynamic limit and the resulting statistical model would have a true spin glass transition. The effective Hamiltonian in this case would, however, have quenched disorder.

To support the spin glass analogy for the case of binary contrast, where the corresponding Hamiltonian has no quenched disorder, 
we performed numerical simulations on systems up to block size $N=23$. The Metropolis algorithm was used to sample the distribution $p(k)$ defined by (\ref{p(k)}), (\ref{p(k|x)}) and (\ref{intens}). In every update the change in each component was limited to $\Delta k_n\in\{-1,0,1\}$. From the average $\langle -\log_2{p(k)}\rangle$ we obtained the entropy $H(K)$. The conditional entropy $H(K|X)$ involved less effort because the sum over counts $k$ can be evaluated explicitly, leaving a single sum over the approximately $M=2^{N-2}/N$ symmetry classes of binary sequences to be performed numerically. By contrast, such a sum needs to be performed for every update in the sampling of $p(k)$. The mutual information is obtained from the difference of entropies, which we now normalize as $I_B(\mu)=I_B(X,K)/\log_2{M}$ to take into account finite size effects arising from the symmetry classes.

The counterpart of the Edwards-Anderson spin glass order parameter \cite{EdwardsAnderson}, for our model, is given by
\begin{equation}
q(\mu)=\frac{1}{N-1}\sum_{n=1}^{N-1}\left(\langle k_n(t)\rangle_t/\mu-1\right)^2,
\end{equation}
where a time average is taken over the photon counts $k(t)$ generated by local $k$-step dynamics. In the ergodic phase, where $\langle k_n(t)\rangle_t=\mu$ is independent of $n$ and $q$ vanishes, the time series origin need not be specified. But because $k(0)$ does matter in the spin glass phase, we additionally average $q(\mu)$ with respect to $k(0)$ in all the simulations. In this ``basin average", we sample $k(0)$ by uniformly sampling the binary sequence intensities $w(x)$, $x\in B$, and setting $k_n(0)=\lfloor w_n\rfloor$ (\textit{i.e.} the most likely photon counts for that sequence). Assuming, in the spin glass phase, that the dynamics is restricted to the basin specified by a single intensity distribution $w$, we have $\langle k_n(t)\rangle_t=w_n$ and the basin average, indicated by the overbar, takes the form
\begin{equation}\label{qave}
\bar{q}(\mu)=\frac{1}{N-1}\sum_{n=1}^{N-1}\left( \overline{w_n^2}/\mu^2-2\,\overline{w_n}/\mu+1\right).
\end{equation}
In the limit of large $N$ the intensities (for $n\ne 0, N/2$) have Wilson statistics with $\overline{w_n}=\mu$ and $\overline{w_n^2}=2\mu^2$, giving the basin average $\bar{q}(\mu)=1$ in the spin glass phase. The basin averages in our simulations used 200 samples.

\begin{figure}[t]
\begin{center}
\includegraphics{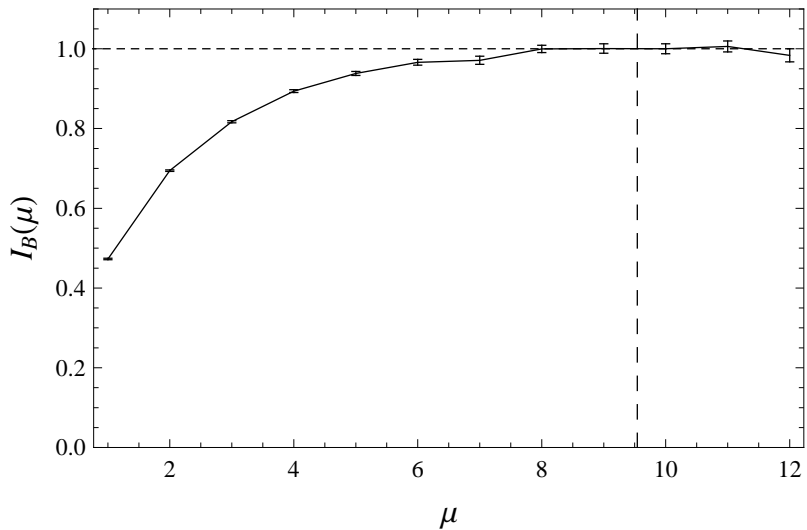}
\end{center}
\caption{Normalized mutual information, computed numerically for binary codewords of length $N=23$, as a function of the mean photon number per bit, $\mu$. The function $I_B(\mu)$ would equal unity for $\mu>\mu_\mathrm{c}\approx 9.543$ (dashed line) in the case of a random Gaussian code. For binary codewords, however, the saturation is imperfect by an exponentially decaying function of $\mu$ that is not resolved by the computation.}
\end{figure}

\begin{figure}[t]
\begin{center}
\includegraphics{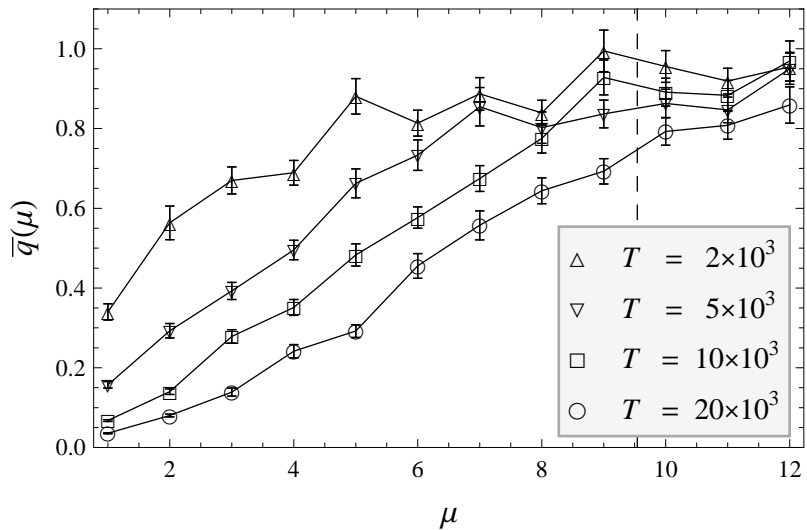}
\end{center}
\caption{Counterpart of the Edwards-Anderson order parameter $\bar{q}(\mu)$, computed numerically for the binary codeword spin glass model with $N=23$ as a function of $\mu$. The averaging time $T$ (number of Metropolis updates) for ergodic behavior grows dramatically as $\mu$ approaches the transition region at $\mu_c\approx 9.5$.}
\end{figure}

Simulation results for the mutual information $I_B(\mu)$ and the order parameter $\bar{q}(\mu)$ are shown in Figures 4 and 5. We see that $I_B(\mu)$ saturates quickly to 1 for $\mu$ near the value $\mu_c\approx 10$ where the FPC has 1 bit capacity. From our analysis of decoding error rates in the previous section, however, we expect that $I_B(\mu)$ fails to saturate by an amount that is exponentially decreasing in $\mu$ even at infinite $N$. Our simulation data show that these corrections are actually quite negligible. The spin glass quasi-transition, in the case of $\bar{q}(\mu)$, is most evident in the behavior with respect to the averaging time period. We see in Figure 5 that as $\mu_c$ is approached from below, the vanishing of the order parameter, and the restoration of ergodicity, requires increasingly long averaging times. For $\mu>\mu_c$ the kinetics of basin-hoping is likely dominated by pairs of sub-distributions $p(k|x)$ with sequences $x$ differing by a single flipped bit. Since hopping to a new sub-distribution is similar to making a decoding error, we expect the time scale for ergodic behavior to grow exponentially with $\mu$ in this (quasi) spin glass ordered phase.

\section{Performance of iterative phase retrieval algorithms}

Practical phase retrieval algorithms, for reconstructing contrast from intensity data, are generally not designed to deal with noise. Maximum likelihood decoding (section \ref{Fidelity}), although optimal from the perspective of noise, is not feasible as it requires an exhaustive search. For the binary contrast problem we have seen that unique (zero flipped bit) reconstructions are impossible for $\mu<\mu_c\approx 9.543$ and that above $\mu_c$ the error rate decays so rapidly that essentially perfect reconstructions are possible in principle. The behavior of practical phase retrieval algorithms in this transition regime is therefore of interest. We have performed extensive simulations with the difference map algorithm \cite{diffmapPNAS} that show its ability to deal with noise is close to that of the best possible algorithm.
Because the simulations in the present study used exactly the same constraint projections and parameters given previously \cite{diffmapPNAS}, we only need to specify how the algorithm was adapted to work with noisy data.

\begin{figure}[t!]
\begin{center}
\includegraphics{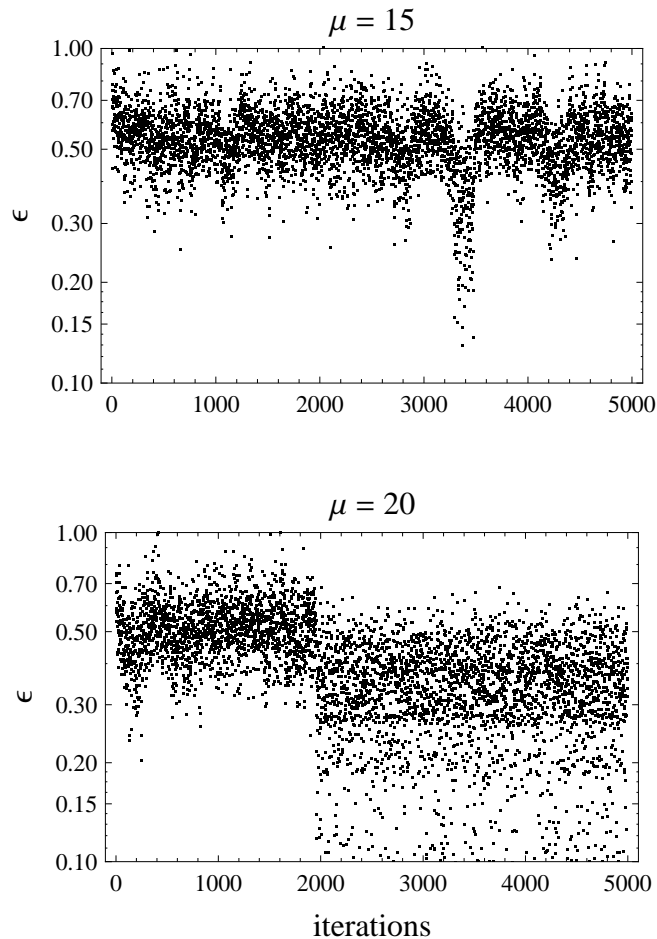}
\end{center}
\caption{Time series of the difference map error $\epsilon$ for two values of the noise parameter and $N=50$; top: $\mu=15$, bottom: $\mu=20$. Incompatibility of constraints as a result of noise prevents $\epsilon$ from vanishing. As the uniqueness threshold $\mu\approx 10$ is approached, the two-state distribution seen in the lower panel tends to the single steady-state of the upper panel. The reconstructions obtained at the lowest $\epsilon$ in both of these experiments yielded valid solutions.}
\end{figure}

\begin{figure}[t]
\begin{center}
\includegraphics{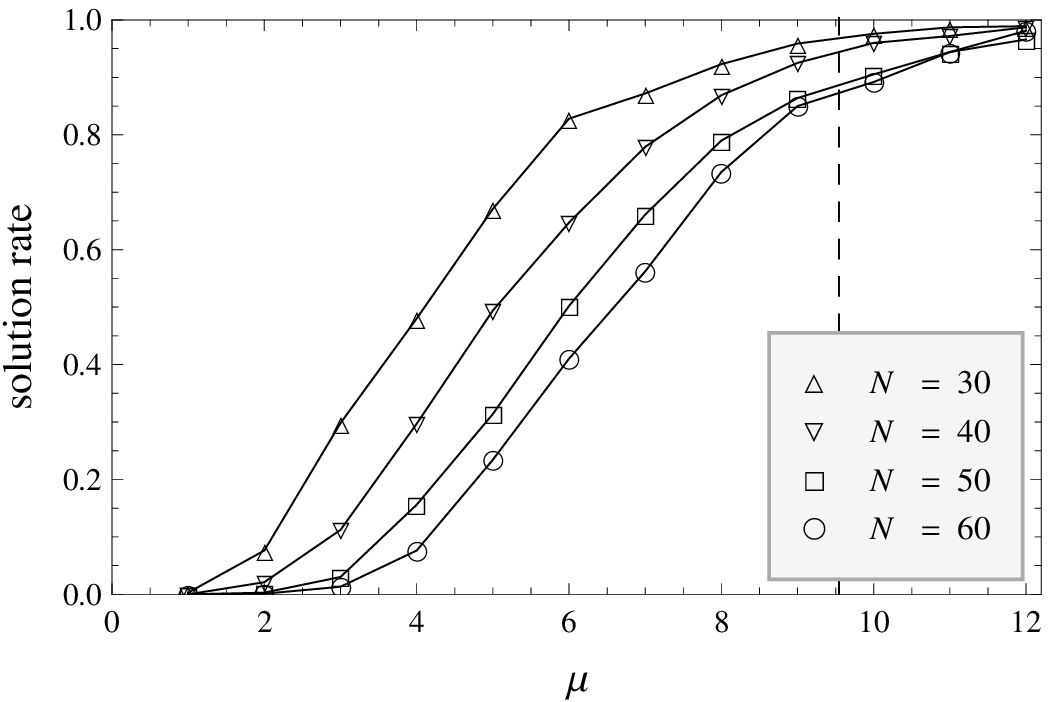}
\end{center}
\caption{Phase retrieval success rate of the difference map algorithm as a function of the mean photon number per bit, $\mu$, for four different binary sequence lengths $N$.}
\end{figure}

The input for each phase retrieval experiment was a single Poisson sampling, $k_n$, of the true intensities $w_n$ of a binary sequence selected at random. Given a photon count $k_n$ the most-probable intensity estimate, based only on Poisson statistics, is $\tilde{w}_n=k_n/(1+1/\mu)$. Although this is simply a uniform rescaling, it needs to be applied in order to preserve the $\pm 1$ scale of the binary contrast projection used by the algorithm. After symmetrization of Friedel pairs, these intensity values were used as hard constraints in the algorithm's Fourier magnitude constraint projection. Because of the errors in the magnitude estimates, the constraint satisfaction problem no longer has a true solution, and a protocol must be established to recover binary sequence candidates in the absence of fixed points. Our procedure was to let the algorithm run for a fixed number of iterations and keep a record of the smallest error metric $\epsilon$ over the course of the run. The binary sequence returned by the binary-value projection, at the iteration with smallest $\epsilon$, was then output as the solution candidate. By repeating such experiments, each derived from a different, randomly selected sequence, we compiled reconstruction success rates for various values of $\mu$. A reconstruction was considered a success if the Fourier intensities of the solution candidate exactly matched those of the sequence that produced the noisy data. Samples of the $\epsilon$ time series for successful reconstructions at two levels of noise are compared in Figure 6.

Plots of the reconstruction success rates are shown in Figure 7 for a range of sequence lengths up to $N=60$. For each $N$, the same number of iterations was used for each value of $\mu$, and this number was increased until there was no change in the success rates, as measured by the mean of $10^3$ trials. At the largest size, $N=60$, each solution attempt used $10^7$ iterations.
These simulations show that the success rate of the difference map algorithm is close to the best possible. The behavior of the success rates with respect to $N$, specifically the narrowing of the transition region, is consistent with the existence of a thermodynamic transition.

\section{Summary and conclusions}

We have shown that in the case of binary contrast reconstruction, where the free variables are not continuous, and in the limit of strong photon shot noise, such that the intensities do not provide precisely known data, an information theoretic framework is useful and should replace constraint counting as the means for assessing the feasibility of phase retrieval. The usual criterion of a reconstruction problem being under- or over-constrained is replaced by a criterion based on the mutual information between noisy data and contrast variables. When the mutual information exceeds the target entropy of the contrast, a unique reconstruction is possible in principle. More generally, the mutual information provides a measure of the information that can be extracted from the intensity measurements regardless of noise.

The mutual information exhibits a transition much like that of a spin glass, from an equilibrium ``paramagnetic" phase at high noise, where very many contrasts are equally compatible with the data, to an ``ordered" phase  at low noise comprising multiple equilibrium states, each associated with a unique pairing of data with contrast. We believe this correspondence extends beyond the minimalist model studied here, and reflects favorably on the robustness and prospects of diffractive imaging in general.

This study has focused on a model with periodic contrast. A result of general interest is the information content $I_G(\mu)$ of an average Bragg peak as a function of the mean photon count $\mu$ when the contrast has only a weak (Gaussian) prior. This information measure was then used to obtain the noise threshold for obtaining just one bit of information per contrast pixel, as would be the case for an image of magnetic domains.

\begin{figure}[t!]
\begin{center}
\includegraphics{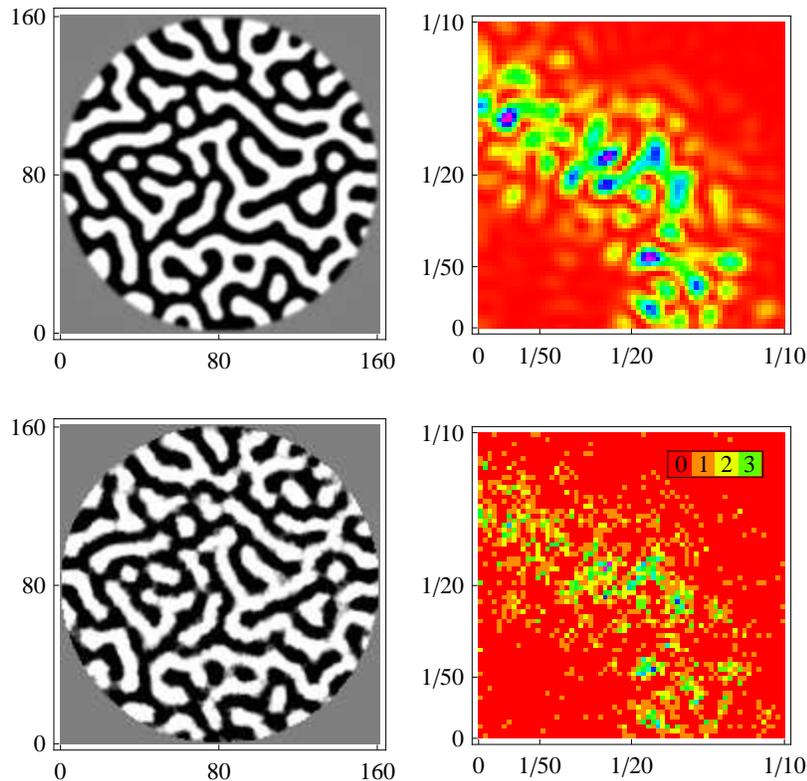}
\end{center}
\caption{Simulations of realistic magnetic domain reconstruction \cite{realistic}. Top row: simulated Ising domain pattern in a circular aperture and one quadrant of its continuous diffraction pattern. Bottom row: domains reconstructed using the difference map algorithm from the noisy data on the right (6047 photons from all four quadrants, or about 0.3 photons per aperture pixel).}
\end{figure}

A natural extension of this work would be the study of non-periodic contrast with a known support. The mutual information would then depend not just on the number of photons detected per pixel of the support, but presumably also on the support shape. Encouraged by the results of this study we have performed simulations of magnetic domain reconstruction within a circular support at very low photon counts. Preliminary results shown in Figure 8 indicate that high fidelity reconstructions in this more realistic setting can be achieved at very high levels of noise. 
 
\ack 
V. E. thanks Jim Sethna for his comments on the manuscript and the Kavli Institute for Theoretical Physics where this work was completed; KITP is supported by the 
National Science Foundation under Grant No. NSF PHY05-51164. Support for this work was provided by the Department of Energy grant DE-FG02-05ER46198.


\section*{References}
\begin{thebibliography}{10}
\bibitem{Millane} Millane R P 1990 {\it J. Opt. Soc. Am.} A {\bf 7} 394-411
\bibitem{Hauptman} Hauptman H A 1991 {\it Rep. Prog. Phys.} {\bf 54} 1427-54
\bibitem{experiment} Gutt C {\it et al.} 2010 {\it Phys. Rev.} B {\bf 81} 100401
\bibitem{Shannon} Shannon C E 1948 {\it Bell System Tech. J.} {\bf 27} 379-423, 623-56
\bibitem{Wilson} Wilson A J C 1949 {\it Acta Cryst.} {\bf 2} 318-21
\bibitem{PCC} Goodwin B E and Bolgiano Jr. L P 1965 {\it Proc. IEEE} {\bf 53} 1745 - 6
\bibitem{SpinGlass} Mezard M, Parisi G and Virasoro M 1988 {\it Spin Glass Theory and Beyond} (World Scientific)
\bibitem{Sourlas} Sourlas N 1989 {\it Nature} {\bf 339} 693-5
\bibitem{MM} M\'ezard M and Montanari A 2009 {\it Information, Physics, and Computation} (Oxford University Press)
\bibitem{EdwardsAnderson} Edwards S F and Anderson P W 1975 {\it J. Phys. F: Metal Phys.} {\bf 5} 965-74
\bibitem{diffmapPNAS} Elser V, Rankenburg I and Thibault P 2007 {\it Proc. Natl. Acad. Sci. USA} {\bf 104} 418-23
\bibitem{realistic} Loh N D, Eisebitt S, Flewett S and Elser V 2010 submitted to {\it Phys. Rev.} E
\end{thebibliography}
\end{document}